\documentclass[prd,twocolumn,tightenlines,superscriptaddress,nofootinbib,
showpacs]{revtex4}
\usepackage{amssymb,latexsym}
\usepackage{amsmath,amsbsy,bbm}
\usepackage{epsfig,bm}
\usepackage{graphicx,comment}
\usepackage{color}
\usepackage{soul}
\usepackage[normalem]{ulem}
\begin{document}

\title{Classification for the universal scaling of N\'eel temperature and staggered magnetization density of three-dimensional dimerized spin-1/2 antiferromagnets}

\author{D.-R. Tan}
\affiliation{Department of Physics, National Taiwan Normal University,
88, Sec.4, Ting-Chou Rd., Taipei 116, Taiwan}
\author{C.-D. Li}
\affiliation{Department of Physics, National Taiwan Normal University,
88, Sec.4, Ting-Chou Rd., Taipei 116, Taiwan}
\author{F.-J. Jiang}
\email[]{fjjiang@ntnu.edu.tw}
\affiliation{Department of Physics, National Taiwan Normal University,
88, Sec.4, Ting-Chou Rd., Taipei 116, Taiwan}

\begin{abstract}
Inspired by the recently theoretical development relevant to the experimental 
data of TlCuCl$_3$, particularly those associated with the universal 
scaling between the N\'eel temperature $T_N$ and the staggered magnetization 
density $M_s$, we carry a detailed investigation of 3-dimensional (3D) 
dimerized quantum antiferromagnets using the first principles quantum Monte 
Carlo calculations. The motivation behind our study is to better understand 
the microscopic effects on these scaling relations of $T_N$ and $M_s$, hence 
to shed some light on some of the observed inconsistency between the 
theoretical and the experimental results. Remarkably, for the considered 3D 
dimerized models, we find that the established universal scaling relations 
can indeed be categorized by the amount of stronger antiferromagnetic 
couplings connected to a lattice site. Convincing numerical evidence is
provided to support this conjecture. The relevance of the outcomes
presented here to the experiments of TlCuCl$_3$ is briefly discussed as well. 
         
\end{abstract}


\maketitle

\section{Introduction}

\begin{figure}
\begin{center}
\vbox{
\hbox{
\includegraphics[width=0.23\textwidth]{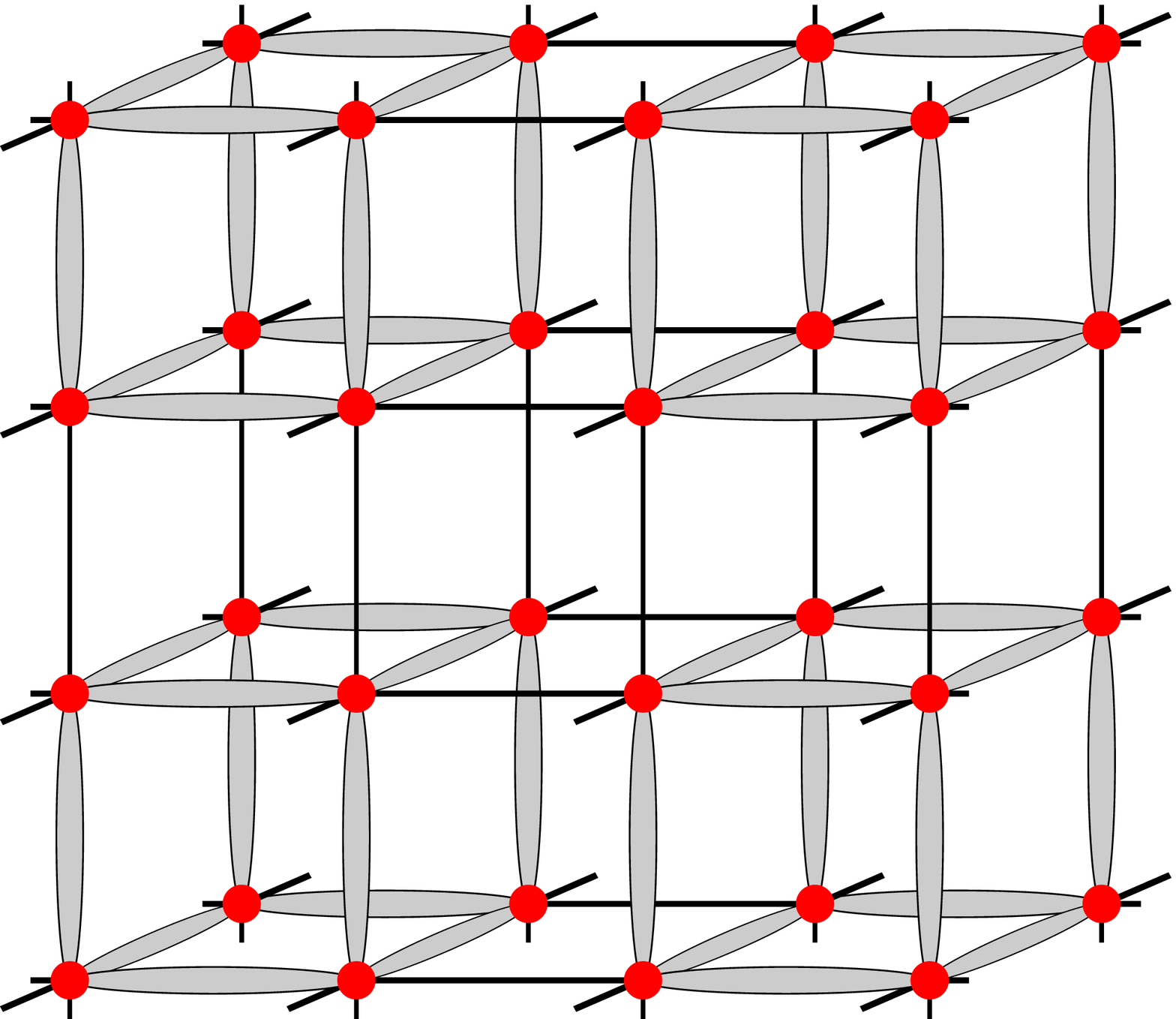}~~~
\includegraphics[width=0.23\textwidth]{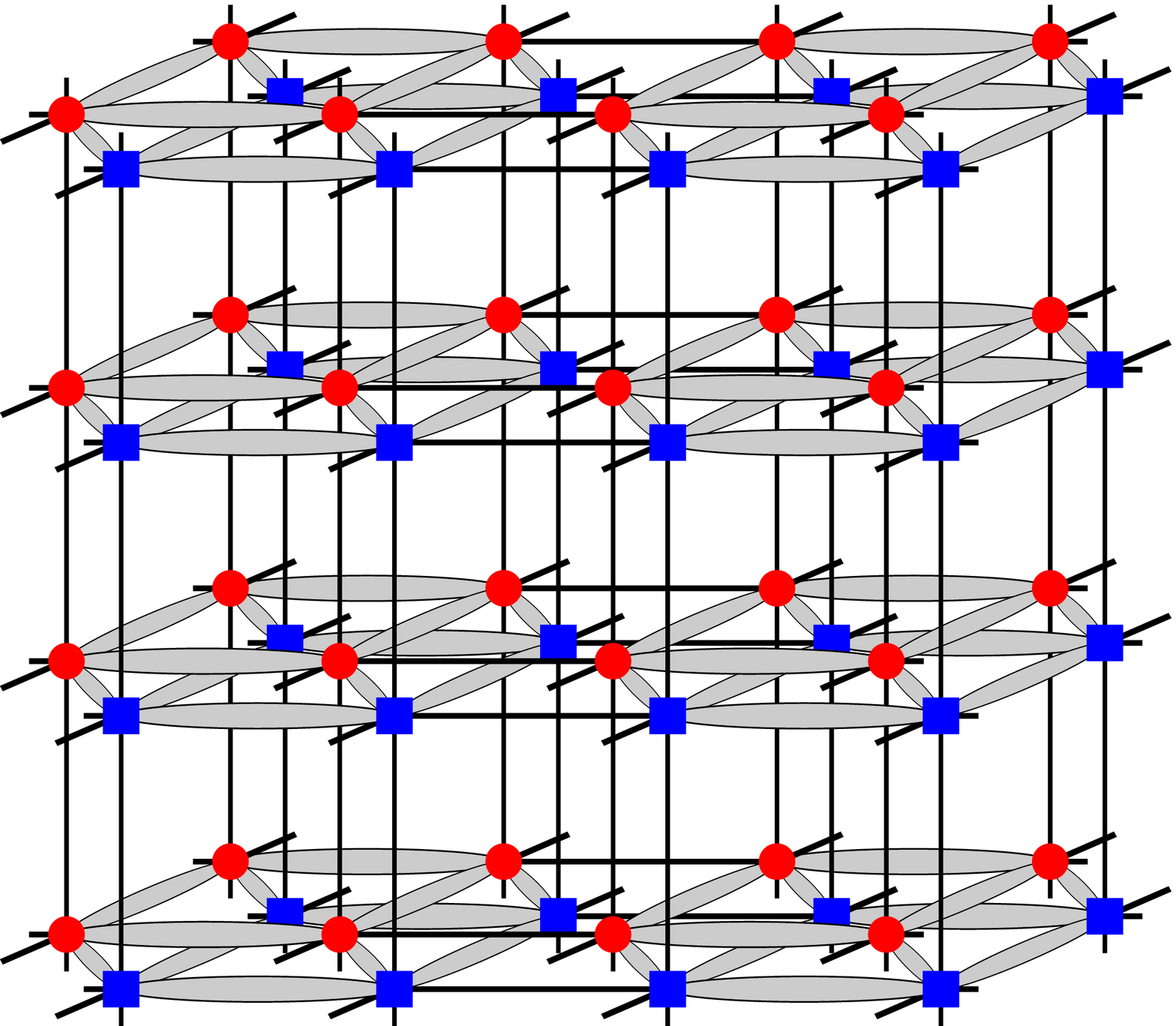}
}\vskip0.25cm
\hbox{
\includegraphics[width=0.23\textwidth]{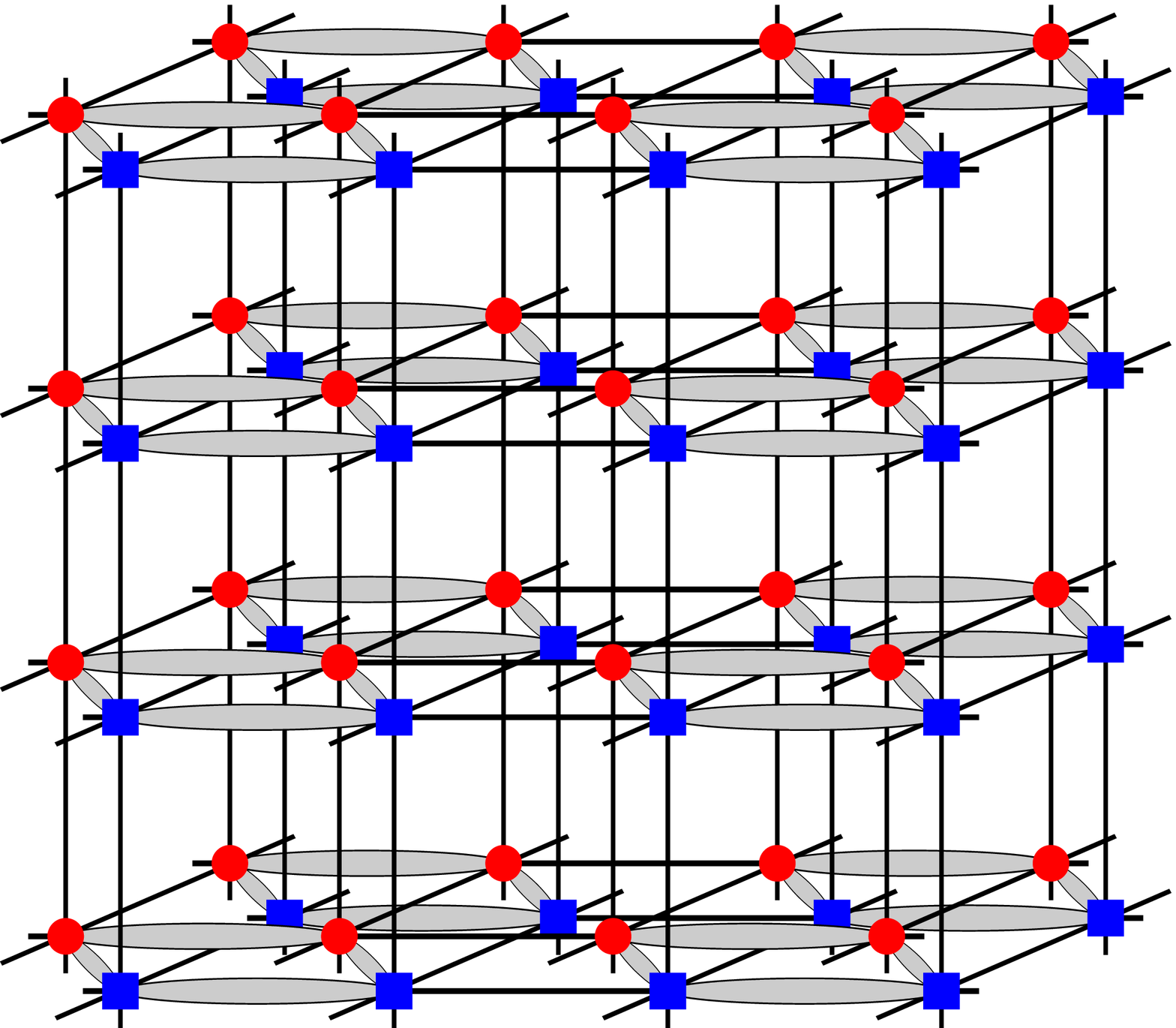}~~~
\includegraphics[width=0.23\textwidth]{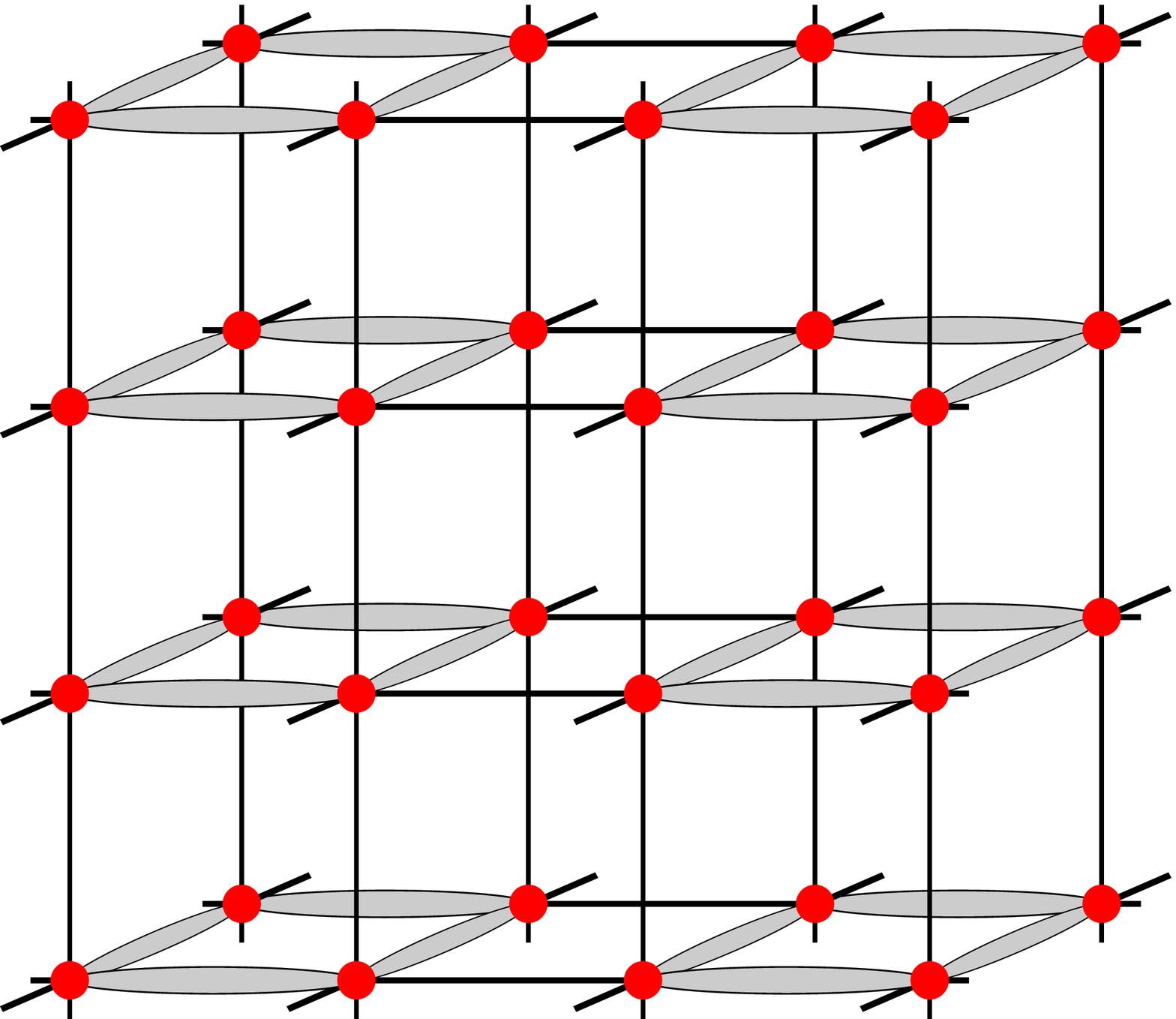}
}}
\end{center}\vskip-0.2cm
\caption{The 3D dimerized spin-1/2 Heisenberg models investigated in
this study: 3D cubical model (top left), double-cube-plaquette model (top right), 
double-cube-ladder model (bottom left), and 3D plaquette model (bottom right).
Notice the antiferromagnetic coupling strength for the thick bonds and thin 
bonds are given by $J'$ and $J$, respectively. }
\label{model_fig1}
\end{figure}

While in general certain intriguing properties related to the phase transitions of
classical models are governed by the thermal fluctuations, many interesting characteristics 
of different phases of quantum systems are triggered by quantum fluctuations at 
zero temperature \cite{Nig92,Lan94,Car10,Sac11,Rei16}. 
In other words, a great deal of attractive phenomena of quantum systems 
are observed at the low temperature regions where quantum fluctuations play the dominated
roles in determining the properties of these systems. Still, 
for quantum systems, thermal fluctuations and the interplay between the effects from finite 
temperatures and zero temperatures may lead to compelling and fascinating 
results.
A noticeable such an example is the quantum critical regime (QCR) associated with 
two-dimensional (2D) antiferromagnets \cite{Chu93,Chu931,Chu94}. 

Theoretically QCR is characterized by the appearance of several universal behavior
among some physical quantities of the underlying 2D spin systems.
In particular, this regime should be 
detectable at finite temperatures. Based on the relevant analytic
calculations, for dimerized Heisenberg models, this regime 
should exist at any values of the tuning parameters associated with spatial 
anisotropy. Interestingly, while numerical studies of these models
indicate the universal behavior associated with QCR can be observed with ease 
at the finite temperature regions above the related 2D quantum critical points (QCPs),
such generic effects seem to disappear, or at least their existence are 
not firmly established yet, when the related calculations are 
carried out relatively away from QCPs \cite{San95,Tro96,Tro97,Tro98,Kim99,Kim00}. To put it in another way, for quantum systems, 
the exotic characteristics of QCR can only be confirmed 
rigorously at the finite temperature regions above the associated QCPs where a dramatic change in the 
ground states occurs due to very strong quantum fluctuations.
Although intuitively the thermal and ground state properties of a quantum 
system may seem to be unrelated to each other, close connections between 
these two categories of properties of that system may still exist.
 
Recently, the experimental results of TlCuCl$_3$ have stimulated several
theoretical investigation \cite{Rue03,Rue08,Kul11,Oit12,Jin12,Kao13,Mer14,Yan15,Har15,Har17,Har171}. 
In particular, the phase diagram of
TlCuCl$_3$ under pressure motivates a few analytic and numerical explorations of 
three universal scalings between a thermal and a ground state property of
three-dimensional (3D) dimerized quantum
antiferromagnets. Specifically, it is demonstrated that for three different
3D dimerized spin-1/2 Heisenberg models, the data collapse
of the physical quantity $T_N/T^{\star}$ as functions of $M_s$ 
leads to a universal curve \cite{Jin12}. In other words, for these three 
various dimerized systems, when the data of $T_N/T^{\star}$  
are treated as functions of $M_s$, they fall on top of a smooth curve.    
Here $T_N$ is the N\'eel temperature, $T^{\star}$ is the temperature where
the observable uniform susceptibility $\chi_u$ reaches its maximum value,
and $M_s$ is the staggered magnetization density. Similar smooth scaling
appears as well if the quantity $T_N/\overline{J}$ is considered instead
of $T_N/\overline{J}$ \cite{Jin12}. Here $\overline{J}$ is the summation of the 
antiferromagnetic
coupling strength connected to a site of any of the studied dimerized models.
Later it is shown that these scaling relations emerge as well for disordered 
systems \cite{Tan15,Tan17}.

Although the agreement between the data of TlCuCl$_3$ and the related analytic and
numerical results is impressive, some controversial observations need to
be clarified. For instance, while theoretically the appearance of
smooth curves resulting from data collapse seems to support the scenario
that generic scaling relations between $T_N$ and $M_s$ do exist, experimental data 
indicate these universal relations may depends on the microscopic details of
the investigated models \cite{Rue03,Rue08,Mer14}. 

To uncover whether there indeed are generic scaling relations between $T_N$ and 
$M_s$ for 3D dimerized spin-1/2 antiferromagnets, in this study we conduct a
large scale quantum Monte Carlo (QMC) calculation for several 3D spatially anisotropic 
spin-1/2 Heisenberg models. It is interesting to note the models studied in 
Ref.~\cite{Jin12} that lead to universal data collapse have the following property. 
Specifically, among the antiferromagnetic bonds connected to a site, only one bond is 
of stronger coupling strength. Inspired by this observation, the considered 3D dimerized 
systems in this investigation can be classified by the amount of 
strong bonds touching a lattice site.

As anticipated, based on our numerical results, we find the established 
universal scaling relations mentioned above do appear for the models considered here.
While the emergence of such scaling relations is foreseen, 
it is remarkable and unexpected that the data collapse using the related physical 
quantities of models having the same amount of strong
bonds at each lattice site form their individual smooth universal curves. In particular,
the universal scaling curves for models having different number of strong bonds per site
differ from each other.
In other words, the universal scaling considered in this study can be categorized 
by the amount of strong bonds connected to a lattice site.

The detailed investigation presented in this study not only reinforces
the robustness of the known universal scaling between $T_N$ and $M_s$
for 3D dimerized quantum antiferromagnets, our results take these relations 
further by establishing quantitatively the classification of these
relations. We would like to emaphsize the fact that the outcomes
shown here are useful for related experiments as well. For example, 
by comparing the theoretical predictions and the associated data, 
one can propose the most applicable model for the targeted material.
Moreover, this model can then be 
considered to explore some further theoretical properties of that
material.        

The rest of this paper is organized as follows. After the introduction, 
the studied 3D dimerized spin-1/2 models and the measured observables are 
briefly described. Then the obtained numerical data and the resulting analysis 
outcomes are summarized. In particular, the evidence to support the conjecture
regarding the classification of the considered universal scaling relations 
outlined above is discussed in detail. Finally, a section is devoted 
to conclude the investigation presented here.

\section{Microscopic model and observables}
The 3D dimerized quantum Heisenberg models investigated here are 
given by the Hamilton operators
\begin{eqnarray}
\label{hamilton}
H_1 &=& \sum_{\langle ij \rangle}J_{ij}\,\vec S_i \cdot \vec S_{j} 
+ \sum_{\langle i'j' \rangle}J'_{i'j'}\,\vec S_{i'} \cdot \vec S_{j'}, \\
H_2 &=& \sum_{i} J_{\perp}\vec S_{i,1} \cdot \vec S_{i,2} + \sum_{\langle ij \rangle,\alpha=1,2} J_{ij,\alpha}\,\vec S_{i,\alpha} \cdot 
\vec S_{j,\alpha} \nonumber \\ 
&+&\sum_{\langle i'j' \rangle,\alpha=1,2}J'_{i'j',\alpha}\,\vec S_{i',\alpha} 
\cdot \vec S_{j',\alpha},   
\end{eqnarray}
where in Eq.~(1) $J_{ij}$ and $J'_{i'j'}$ are the antiferromagnetic 
couplings (bonds) connecting nearest neighbor spins $\langle  ij \rangle$ 
and $\langle  i'j' \rangle$ located at a 3D cubical lattice, respectively, 
and $\vec{S}_i$ is the spin-1/2 operator 
at site $i$. Notice the $\alpha$ in the second equation,
which takes the value of either 1 or 2, stands for the 
indices of the considered two copies of 3D cubical lattices. 
In addition, the $J_{\perp}$ appearing above are the couplings connecting 
spins that belong to different copies of the two targeted 3D cubical 
lattices. Finally, the other parameters and the operators showing up in Eq.~(2) have the same
definitions as their counterparts without the subscript $\alpha$ in Eq.~(1).  
It should be pointed out that in this study, we have set $J_{ij}$ = $J_{ij,1}$ = $J_{ij,2}$ = $J$
and $J'_{i'j'}$ = $J'_{i',j',1}$ = $J'_{i'j',2}$ = $J_{\perp}$ = $J'$ with $J'> J$ 
for any $\langle ij \rangle$ and $\langle i'j'\rangle$. 
Figure~\ref{model_fig1} demonstrates the four dimerized spin-1/2 models studied here. 
Notice for the models of the top (bottom) two panels in fig.~\ref{model_fig1}, 
among the bonds touching each lattice site, three (two) of them have larger 
magnitude in antiferromagnetic strength than the others. 
For convenience, in this investigation the models in fig.~\ref{model_fig1}  
will be called 3D cubical model (top left), double-cube-plaquette 
model (top right), double-cube-ladder model (bottom left), and 3D plaquette 
model (bottom right), respectively. 
Finally, since the couplings $J'$ and $J$ satisfy $J' > J$, each of the investigated system will
undergo a quantum phase transition when the corresponding ratio $J'/J$ exceeds a particular value.

To determine the N\'eel temperature $T_N$, the staggered magnetization density $M_s$,
as well as $T^{\star}$ of the considered dimerized models, 
the observables staggered structure factor $S(\pi,\pi,L_1,L_2,L_3)$ on a finite lattice
with linear sizes $L_1$, $L_2$, and $L_3$ are measured. In addition, 
both the spatial and temporal winding numbers squared 
($\langle W_i^2 \rangle$ for $i \in \{1,2,3\}$ and
$\langle W_t^2 \rangle$), spin stiffness $\rho_s$,
first Binder ratio $Q_1$, and second Binder ratio $Q_2$ 
are calculated in our simulations as well. 
The quantity $S(\pi,\pi,L_1,L_2,L_3)$ takes the form 
\begin{equation}
S(\pi,\pi,L_1,L_2,L_3) = 3 \langle ( m_s^z )^2\rangle,
\end{equation}
where 
$m_s^z = \frac{1}{L_1L_2L_3}\sum_{i}(-1)^{i_1+i_2+i_3}S^z_i$ with $S^{z}_i$ being
the third-component of the spin-1/2 operator $\vec{S}_i$ at site $i$. 
Moreover, the spin stiffness $\rho_s$ has the following expression
\begin{equation}
\rho_s = \frac{1}{3}\sum_{i=1,2,3} \rho_{si} = \frac{1}{3\beta}\sum_{i=1,2,3}\frac{\langle W_i^2 \rangle}{L_i},
\end{equation}
where $\beta$ is the inverse temperature.
Finally the observables $Q_1$ and $Q_2$ are defined by
\begin{equation}
Q_1 = \frac{\langle |m_s^z| \rangle^2 }{\langle (m_s^z)^2\rangle} 
\end{equation}
and
\begin{equation}
Q_2 = \frac{\langle (m_s^z)^2 \rangle^2}{\langle (m_s^z)^4\rangle},
\end{equation}
respectively. With these observables, the physical quantities required for our 
study, namely $T_N$, $M_s$, and $T^{\star}$, can be calculated accurately.

\section{The numerical results}
To understand the robustness of the scaling relations associated with 
$T_N$ and $M_s$, namely 
to uncover the rules of under what conditions the data collapse employing results from different models 
will lead to the same universal curves, we have carried out a large-scale QMC simulation 
using the stochastic series expansion (SSE) algorithm with very efficient 
loop-operator update \cite{San99}. 
Before presenting the numerical outcomes obtained from
the QMC simulations, it should be pointed out that in our calculations related to 
the double-cube-plaquette model (double-cube-ladder model), due to the spatial arrangement of its 
antiferromagnetic bonds, the linear box sizes (size) $L_1$ and $L_2$ ($L_1$) used in the 
simulations are twice that of $L_3$ (those of $L_2$ and $L_3$) for most of the considered $J'/J$ 
($J'/J \ge 4.6$) \cite{note1}. This strategy guarantees the aspect ratios among the three spatial winding numbers
squared are kept within certain range. Consequently the 3D features of these models are
preserved. For the 3D cubical model and the 3D plaquette model, the condition $L_1$ = $L_2$ = $L_3$ is 
used in the related calculations.  

In the following, we will firstly detail the determination of $M_s$.

\begin{figure}
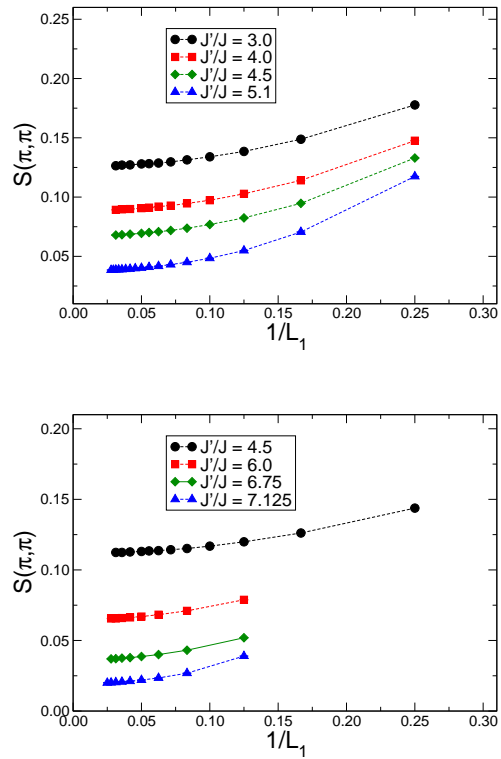

\begin{center}
\vbox
{\includegraphics[width=0.36\textwidth]{Spipi_L_cube.eps}\vskip0.8cm
\includegraphics[width=0.36\textwidth]{Spipi_L_double_plaq.eps}
}
\end{center}\vskip-0.2cm
\caption{The $1/L_1$ dependence of the staggered structure factors 
$S(\pi,\pi)$ for several considered $J'/J$ of
the 3D cubical model (top panel) and the double-cube-plaquette model 
(bottom panel). 
The dashed lines are added to guide the eye.}
\label{ms_fig1}
\end{figure}

\subsection{The determination of $M_s$}
The observable considered for the calculations of 
$M_s$ is $S(\pi,\pi)(L_1)$ \cite{note2}. Specifically, 
for a given $J'/J$, the associated $M_s$ 
is given by $\sqrt{S(\pi,\pi)(L_1 \rightarrow \infty)}$. 
We would like to point out that to determine $M_s$ using this approach, 
the zero temperature, namely the ground state values of $S(\pi,\pi)(L_1)$ are required.
Therefore the simulations related to the calculations of $M_s$ are conducted
using the condition $\beta = 2L_1$ \cite{beta}. For each of the considered models,
we have additionally carried out several simulations (for some selected $J'/J$) 
with $\beta = 4L_1$. The results obtained from these trial calculations agree 
very well with those determined by employing the relation $\beta = 2L_1$ in the 
simulations.  

\begin{figure}
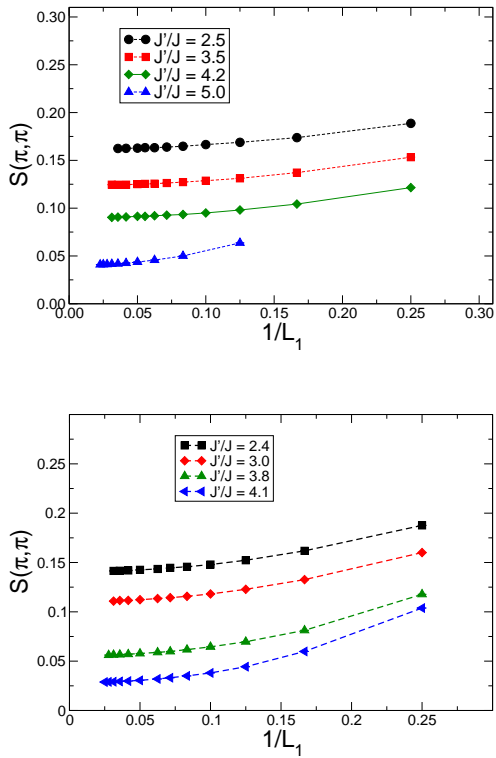

\begin{center}
\vbox
{
\includegraphics[width=0.36\textwidth]{Spipi_L_double_ladder.eps}\vskip0.8cm
\includegraphics[width=0.36\textwidth]{Spipi_L_plaq.eps}
}
\end{center}\vskip-0.2cm
\caption{The $1/L_1$ dependence of the staggered structure factors 
$S(\pi,\pi)$ for several considered $J'/J$ of  
the double-cube-ladder model (top panel)
and the 3D plaquette model (bottom panel). 
The dashed lines are added to guide the eye.}
\label{ms_fig2}
\end{figure}

\begin{figure}
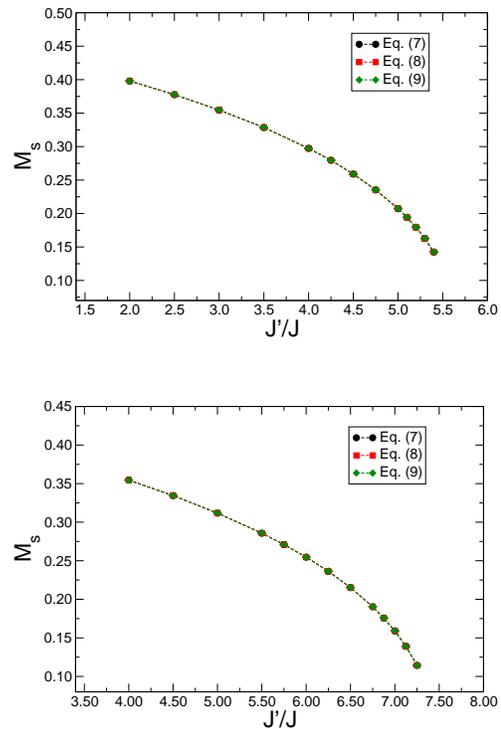

\begin{center}
\vbox{
\includegraphics[width=0.36\textwidth]{Ms_cube.eps}\vskip0.8cm
\includegraphics[width=0.36\textwidth]{Ms_double_plaq.eps}
}
\end{center}
\vskip-0.3cm
\caption{$M_s$ as functions of the considered $J'/J$ for the 
3D cubical model (top panel) and the double-cube-plaquette model 
(bottom panel). The dashed lines are added to guide the eye.}
\label{ms_fig3}
\end{figure}

\begin{figure}
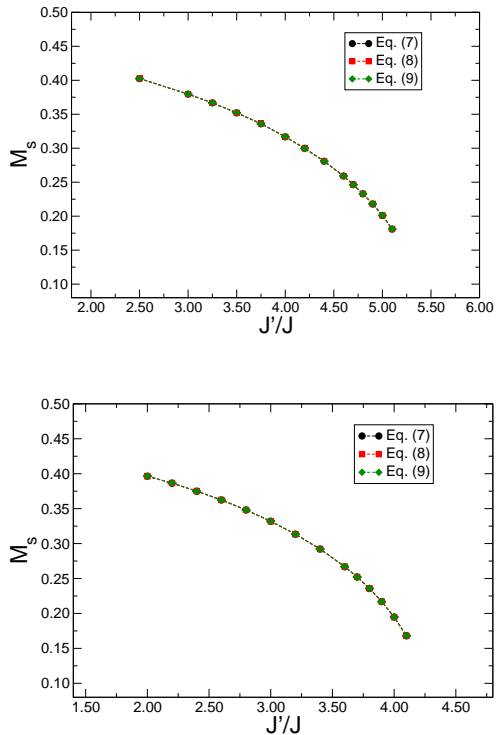

\vskip0.25cm
\begin{center}
\vbox{
\includegraphics[width=0.36\textwidth]{Ms_double_ladder.eps}\vskip0.8cm
\includegraphics[width=0.36\textwidth]{Ms_plaq.eps}
}
\end{center}
\vskip-0.2cm
\caption{$M_s$ as functions of the considered $J'/J$ for 
the double-cube-ladder model (top panel) and the 3D plaquette model 
(bottom panel). 
The dashed lines are added to guide the eye.}
\label{ms_fig4}
\end{figure}

For each of the studied model, the corresponding $1/L_1$-dependence of the ground 
state $S(\pi,\pi)$ for some considered $J'/J$ is depicted 
in figs.~\ref{ms_fig1} and \ref{ms_fig2}. Motivated by the theoretical predictions 
in Ref.~\cite{Car96}, 
the determination of $M_s$ is done by 
extrapolating the staggered structure factors at finite box sizes 
to their bulk results, using the following three ansatzes
\begin{eqnarray}
\label{poly2}&&a_0 + a_2/L^2, \\
\label{poly3}&&b_0 + b_2/L^2 + b_3/L^3, \\
\label{poly4}&&c_0 + c_2/L^2 + c_3/L^3 + c_4/L^4.
\end{eqnarray}
For each good fit ($\chi^2/{\text{DOF}} \le 2.0$), the corresponding 
bulk $M_s$ is calculated by $M_s = \sqrt{F}$ with $F = a_0,b_0,$
or $c_0$ depending on which ansatz is used for the fit. 
The numerical values of $M_s$ 
determined from the fits employing ansatzes (\ref{poly2}), (\ref{poly3}),
and (\ref{poly4}) for all the four models are shown in figs.~\ref{ms_fig3} and \ref{ms_fig4}. 
The agreement between the results of $M_s$ determined from different ansatzes is
remarkably good, and the ones obtained with ansatz (\ref{poly3}) are
used in the following analysis.

We would like to emphasize the fact that since three spatial dimensions 
is the upper critical dimension of the quantum 
phase transitions investigated in this study, when approaching the critical points
one expects to observe logarithmic corrections to $M_s$ (and $T_N$ as well). 
The theoretical calculations of the critical exponents associated with these
logarithmic corrections are available in Refs.~\cite{Ken04,Ken12,Yan15}, and 
the predicted values are confirmed by careful analyses of $M_s$ and $T_N/\overline{J}$ conducted in 
Refs.~\cite{Yan15,Tan17}. To perform an analysis associated with the mentioned logarithmic 
corrections requires data of $M_s$ close to the related quantum critical points. Besides, the motivation of the 
investigation presented here is to understand to what extent the considered scaling relations 
are universal. Therefore, a detailed exploration of the
logarithmic corrections related to the investigated phase transitions will be left for
a future project.

\subsection{The determination of $T_N$}

\begin{figure}
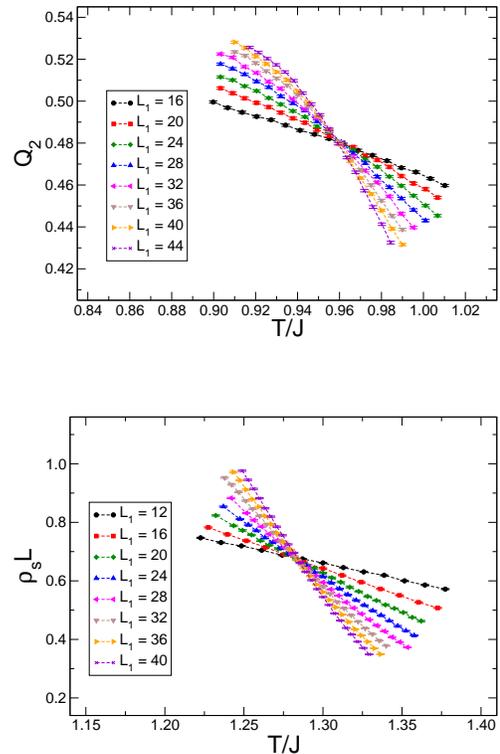

\vskip0.5cm
\begin{center}
\vbox{
\includegraphics[width=0.36\textwidth]{regular_cube_Q2J5.0.eps}\vskip1cm
\includegraphics[width=0.36\textwidth]{double_plaq_rhosLJ6.5.eps}
}
\end{center}
\caption{Top panel: $Q_2$ of the 3D cubical model as functions of $T/J$
for $J'/J = 5.0$ and $L_1$ = 16, 20, 24, 28, 32, 36, 40, 44. 
Bottom panel: $\rho_sL$ of the double-cube-plaquette model as functions of $T/J$
for $J'/J = 6.5$ and $L_1$ = 12, 16, 20, 24, 28, 32, 36, 40.
$J$ is 1.0 in our calculations. The dashed lines are added to guide the eye.}
\label{TN_fig1}
\end{figure}

\begin{figure}
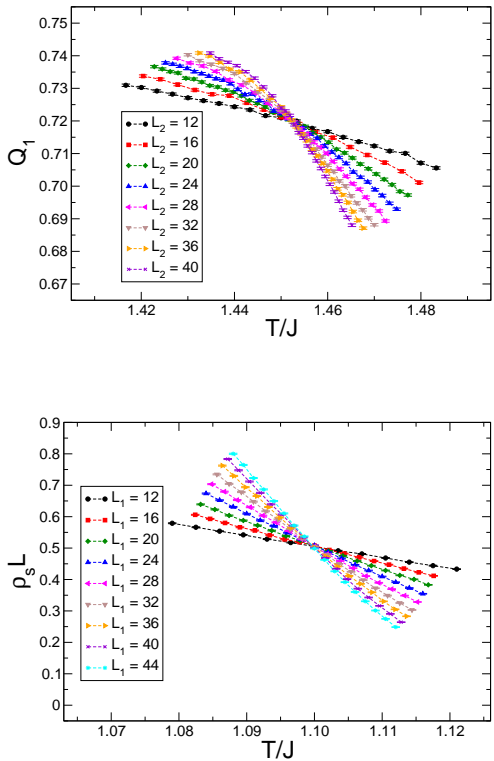

\vskip0.5cm
\begin{center}
\vbox{
\includegraphics[width=0.36\textwidth]{double_ladder_Q1J3.5.eps}\vskip1cm
\includegraphics[width=0.36\textwidth]{regular_plaq_rhosLJ3.0.eps}
}
\end{center}
\caption{Top panel: $Q_1$ of the double-cube-ladder model as functions of $T/J$
for $J'/J = 3.5$ and $L_2$ = 12, 16, 20, 24, 28, 32, 36, 40. 
Bottom panel: $\rho_sL$ of the 3D plaquette model as functions of $T/J$
for $J'/J = 3.0$ and $L_1$ = 12, 16, 20, 24, 28, 32, 36, 40, 44.
$J$ is 1.0 in our calculations. The dashed lines are added to guide the eye.}
\label{TN_fig2}
\end{figure}

The N\'eel temperatures $T_N$ for various $J'/J$ of the four studied models are 
calculated from the observables $\rho_sL$ (which is given by
$\left(\sum_{i=1}^3 \rho_{si}L_i\right)/3$), $Q_1$, as well as $Q_2$. 
Notice bootstrap-type fits using constrained standard 
finite-size scaling ansatz of the form 
$(1+b_0L^{-\omega})(b_1 + b_2tL^{1/\nu} +b_3(tL^{1/\nu})^2+...$), up to second,
third, and (or) fourth order in $tL^{1/\nu}$ are performed in the determination
of $T_N$. 
Here $b_i$ for $i=0,1,2,...$ are some constants and $t = \frac{T-T_N}{T_N}$. 
For some $J'/J$, ansatz up to fifth order in $tL^{1/\nu}$ is used. 
The data of $\rho_sL$, $Q_1$, and $Q_2$ of some considered $J'/J$ for the
investigated models are shown in figs.~\ref{TN_fig1} and \ref{TN_fig2}. 

\begin{figure}
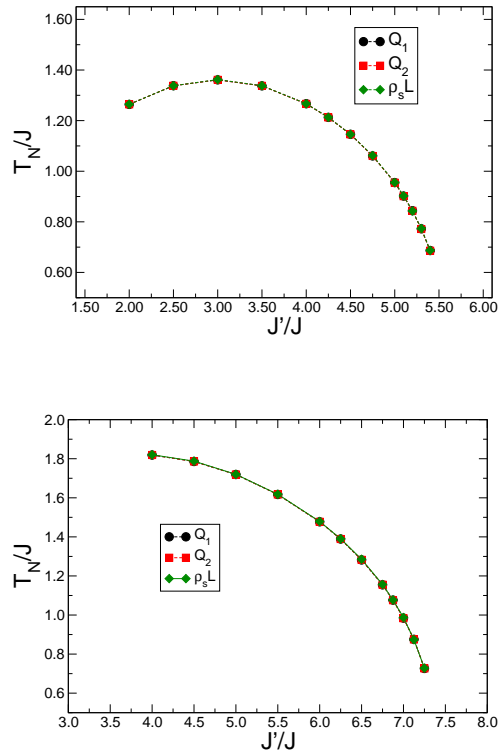

\vskip0.5cm
\begin{center}
\vbox{
\includegraphics[width=0.36\textwidth]{regular_cube_TN.eps}\vskip1cm
\includegraphics[width=0.36\textwidth]{double_plaq_TN.eps}
}
\end{center}
\caption{The $J'/J$ dependence of $T_N$ obtained from $Q_1$, $Q_2$,
and $\rho_sL$ for the 3D cubical model (top panel) and 
the double-cube-plaquette model (bottom panel), respectively.
$J$ is 1.0 in our simulations.}
\label{TN_fig3}
\end{figure}

\begin{figure}
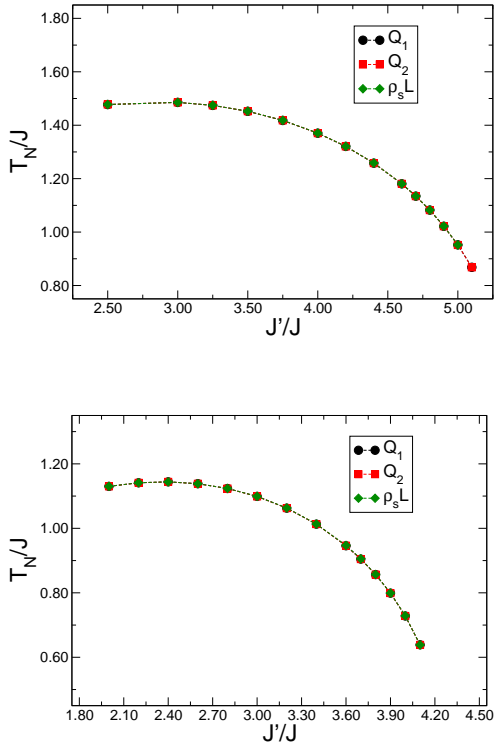

\vskip0.5cm
\begin{center}
\vbox{
\includegraphics[width=0.36\textwidth]{double_ladder_TN.eps}\vskip1cm
\includegraphics[width=0.36\textwidth]{plaq_TN.eps}
}
\end{center}
\caption{The $J'/J$ dependence of $T_N$ obtained from $Q_1$, $Q_2$,
and $\rho_sL$ for the double-cube-ladder model (top panel) and the
3D plaquette model (bottom panel). $J$ is 1.0 in our simulations.
Notice the $T_N$ from $\rho_sL$ for $J'/J$ = 5.1 of the double-cube-ladder 
model is not included in the sub-figure.}
\label{TN_fig4}
\end{figure}

In our analysis related to the calculations of $T_N$, a fit is treated as 
a good fit if the corresponding $\chi^2/{\text{DOF}}$ satisfies 
$\chi^2/{\text{DOF}} \le 2.0$. For few cases, in particular
those associated with the observables $\rho_s L$, the criterion for good fits
is slightly less restricted ($\chi^2/{\text{DOF}} \le 2.5$ is used for these 
situations). For every $J'/J$ of each studied model, fits are carried out with 
ansatzes of various order in $tL^{1/\nu}$. Furthermore, for a given $J'/J$, 
several sets of data 
having different minimum box sizes are considered for the fits as well. 
The quoted values of $T_N$
in this study are estimated by averaging the corresponding results of good fits.
In addition, the error bar of each cited $T_N$ is estimated conservatively from the uncertainty of every 
individual $T_N$ of the associated good fits.
The determined $T_N$ from the three used observables, namely $\rho_s L$,
$Q_1$, and $Q_2$ for all the studied models are shown in figs.~\ref{TN_fig3} and \ref{TN_fig4}.

\subsection{The determination of $T^{\star}$}

\begin{figure}
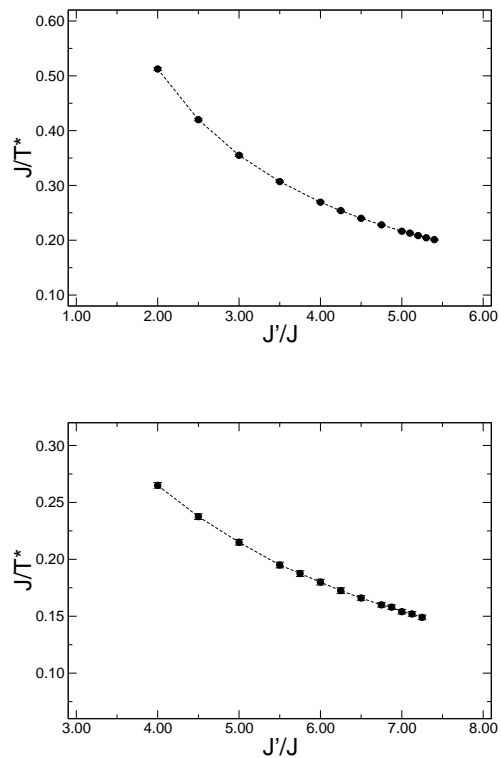

\vskip1.0cm
\begin{center}
\vbox{
\includegraphics[width=0.36\textwidth]{cube_Tstar.eps}\vskip1cm
\includegraphics[width=0.36\textwidth]{double_plaq_Tstar.eps}
}
\end{center}
\caption{The inverse of $T^{\star}$ as functions of $J'/J$ for the 3D cubical model (top panel) and 
the double-cube-plaquette model (bottom panel). $J$ is 1.0 in our simulations.}
\label{Tstar_fig1}
\end{figure}

\begin{figure}
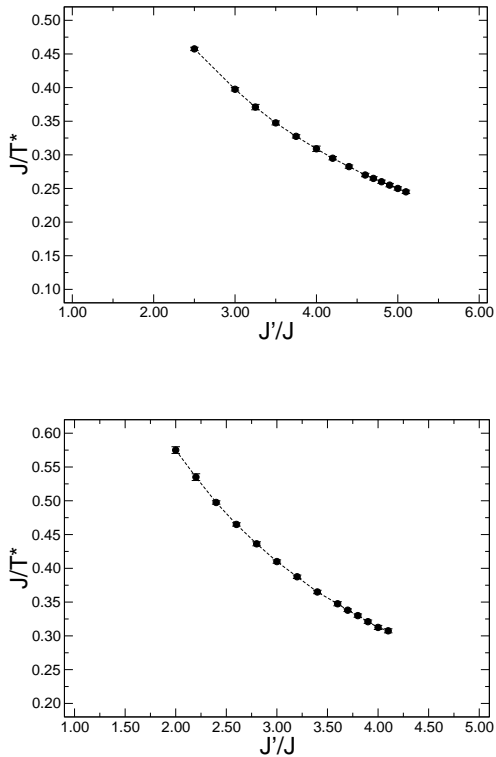

\vskip0.5cm
\begin{center}
\vbox{
\includegraphics[width=0.36\textwidth]{double_ladder_Tstar.eps}\vskip1cm
\includegraphics[width=0.36\textwidth]{plaq_Tstar.eps}
}
\end{center}
\caption{The inverse of $T^{\star}$ as functions of $J'/J$ for the double-cube-ladder model (top panel) and 
the 3D plaquette model (bottom panel). $J$ is 1.0 in our simulations.}
\label{Tstar_fig2}
\end{figure}

For all the four investigated models, the corresponding
$T^{\star}$, namely the temperatures at which $\chi_u$ reach their
maximum value, are determined on lattices with moderate large box sizes
such as $(L_1,L_2,L_3)$ = $(16,16,16)$, $(24,12,12)$, and so on.
The obtained estimations of the inverse of $T^{\star}$ as functions of $J'/J$
are shown in figs.~\ref{Tstar_fig1} and \ref{Tstar_fig2}. 
For each individual model, several additional 
simulations on lattice with larger or smaller box sizes than those associated with
the results demonstrated in figs.~\ref{Tstar_fig1} and \ref{Tstar_fig2} 
are conducted at some selected values of $J'/J$. These trial simulations
confirm that for these selected $J'/J$ the corresponding outcomes presented in 
figs.~\ref{Tstar_fig1} and \ref{Tstar_fig2} are indeed the bulk results.
Therefore the used $T^{\star}$ in the relevant analysis should be
reliable.

\subsection{The scaling relations between $T_N/\overline{J}$, $T_N/T^{\star}$, and 
$M_s$}
\begin{figure}
\vskip0.5cm
\begin{center}
\includegraphics[width=0.36\textwidth]{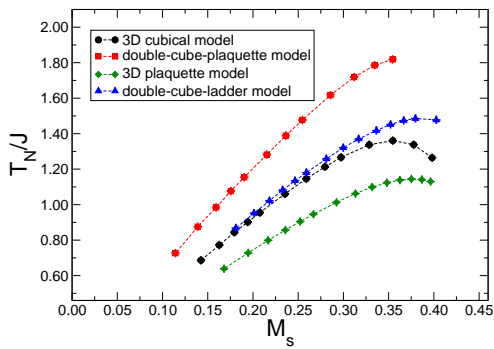}
\end{center}
\caption{$T_N/J$ as functions of $M_s$ for all the considered 3D dimerized models.
The used values of $T_N$ in the figure are from the observable $Q_1$. 
$J$ is set to be 1.0 in our simulations.}
\label{TN_barJ_Tstar_fig1}
\end{figure}

Having obtained $M_s$, $T_N$, and $T^{\star}$, we now turn to study 
the scaling relation(s) between $T_N/\overline{J}$ ($T_N/T^{\star}$) and $M_s$ ($M_s$).
Figure~\ref{TN_barJ_Tstar_fig1} shows $T_N/J$ as functions of $M_s$ for all the
four considered models. The results in fig.~\ref{TN_barJ_Tstar_fig1} indicate 
there is no any universal relations for $T_N/J$ and $M_s$ among the investigated 
dimerized systems. 
  
\begin{figure}
\vskip0.5cm
\begin{center}
\includegraphics[width=0.36\textwidth]{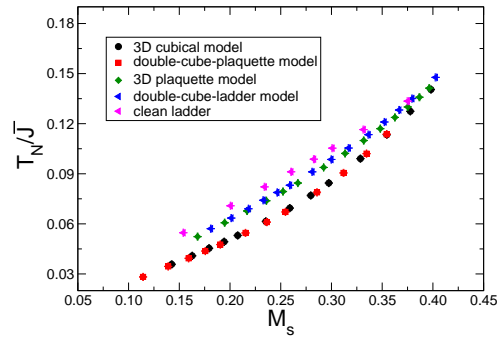}
\end{center}
\caption{$T_N/\overline{J}$ as functions of $M_s$ 
for all the considered models in this study. The used values of $T_N$ in the figure
are from the observable $Q_1$. For comparison purpose,
some results of the 3D dimerized ladder model which has one strong bond per lattice site 
are included here as well \cite{Kao13}.}
\label{TN_barJ_Tstar_fig2}
\end{figure}

\begin{figure}
\vskip0.5cm
\begin{center}
\includegraphics[width=0.36\textwidth]{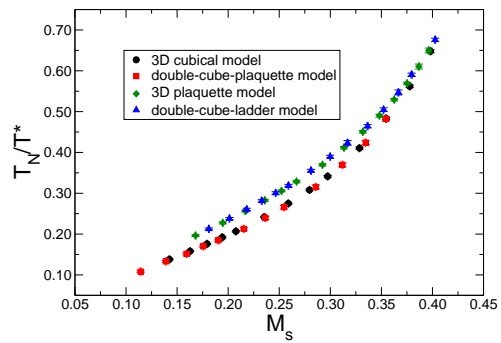}
\end{center}
\caption{$T_N/T^{\star}$ as functions of $M_s$ for all the considered models in this study.
The used values of $T_N$ in the figure are from the observable $Q_1$.}
\label{TN_barJ_Tstar_fig3}
\end{figure}

Remarkably, while no obvious scaling relations are observed 
when $T_N/J$ are treated as functions of $M_s$, such universal dependence
of $T_N$ on $M_s$ do emerge if the quantities $T_N/\overline{J}$
and $T_N/T^{\star}$ are considered. This can be clearly seen in 
figs.~\ref{TN_barJ_Tstar_fig2} and \ref{TN_barJ_Tstar_fig3}.
Specifically, the data of $T_N/\overline{J}$ and $T_NT^{\star}$ of these
studied models do fall on top of their individual universal curves
when these two quantities are regarded as functions of $M_s$. 
The most striking result shown in figs.~\ref{TN_barJ_Tstar_fig2} and 
\ref{TN_barJ_Tstar_fig3} is that these universal scaling curves
can be categorized by the amount of bonds which are connected
to a lattice site and have the stronger antiferromagnetic coupling 
strength $J'$. Indeed,
from the outcomes demonstrated in these figures, one can see that the 
universal curves corresponding to the 3D cubical model and the 
double-cube-plaquette model, which have three bonds 
of coupling strength $J'$ at each of their lattice sites, are different 
from those of the 3D plaquette model and the double-cube-ladder model for which there 
are two bonds of coupling strength $J'$ surrounding every point of their underlying
lattices. Notice for comparison purpose, the data of the 3D dimerized spin-1/2 ladder 
model \cite{Kao13}, which has one strong bond per lattice site, are 
included in fig.~\ref{TN_barJ_Tstar_fig2} as well.

To conclude, figs.~\ref{TN_barJ_Tstar_fig2} and \ref{TN_barJ_Tstar_fig3} show
convincing evidence that the considered universal scaling relations investigated here can be categorized 
by the amount of stronger antiferromagnetic bonds touching any lattice site.
We will argue later that this classification scheme regarding the 
studied universal scaling relations should be a generic one.

\section{Discussions and Conclusions}

For certain types of 3D dimerized quantum antiferromagnets, it is demonstrated
that universal scaling relations appear when the physical quantities 
$T_N/\overline{J}$ and $T_N/T^{\star}$ are considered as functions of $M_s$ \cite{Jin12}. 
Furthermore, near the associated quantum critical points, these mentioned observables
scale linearly with $M_s$. Similar phenomena are observed for disordered models
as well \cite{Tan15}. Motivated by these findings, in this study
we have investigated four 3D dimerized spin-1/2 Heisenberg models, using
the first principles nonperturbative quantum Monte Carlo simulations.
Notice the models studied in Ref.~\cite{Jin12} have the feature that
among the bonds connected to every lattice site there is only one bond 
having stronger antiferromagnetic coupling strength. Based on this observation, 
for the models considered here, either two or three bonds surrounding a lattice site $p$
possess stronger antiferromagnetic coupling strength than the others touching the same site 
$p$.

Remarkably, universal scaling relations associated with $T_N$ and $M_s$ do
emerge for the four models studied here. In particular, among these four 
dimerized systems, the data collapse of $T_N/\overline{J}$ and $T_N/T^{\star}$
of models having the same amount of strong 
bonds at each lattice site do form their individual smooth universal curves. 
Furthermore, the universal scaling
curves of models having two strong bonds at each lattice site are different from
those associated with models possessing three strong bonds per site. 
In other words, the universal scaling considered in this study can be categorized
by the amount of strong bonds connected to a lattice site.
Our findings considerably generalize those established in literature. 
It is interesting to notice the outcomes reached here are consistent with the 
experimental results of TlCuCl$_3$. Indeed the data of TlCuCl$_3$ in Refs.~\cite{Rue03,Rue08,Mer14} 
indicate the curves associated with the universal scaling of $T_N/T^{\star}$ and $M_s$
most likely depend on the microscopic details of the studied systems.
This is in agreement with the main result obtained in our investigation. 

Finally we would like to point out that in Ref.~\cite{Tan17}, it is shown that for both
a 3D spin-1/2 antiferromagnet with the so-called configurational disorder and the 
3D regular dimerized ladder quantum Heisenberg model, data collapse 
of $T_N/\overline{J}$ (as functions of $M_s$) using the results from both systems leads
to a smooth universal curve as well. Notice for a model with 
configurational disorder, each lattice site has exactly one strong bond
for every disordered realization. Furthermore, while the number of bonds
touching every site of the double-cube-type models
considered here is seven, the other two investigated models have six bonds
connecting to any of their lattices. Based on these observations, it is likely that the
results obtained here, namely the considered universal scaling relations
of 3D dimerized spin-1/2 antiferromagnets can be categorized by
the amount of strong bonds touching every lattice site, may be applicable for
disordered systems and other lattice geometries. To verify whether this
is indeed the case or not, simulating 3D antiferromagnets on the honeycomb
lattice and other disordered models will shed some light on justifying this conjecture.


\section{Acknowledgments}
\vskip-0.5cm
This study is partially supported by MOST of Taiwan.


\begin{thebibliography}{99}
\bibitem{note1}
For our calculations of $M_s$ and $T^{\star}$ associated with the double-cube-plaquette model, 
all three spatial linear box sizes are equal for the simulations of $J'/J=4.0$ and 4.5.
\bibitem{note2}
Since for a fixed $J'/J$ of every considered model, the three linear box 
sizes $L_1$, $L_2$ and $L_3$
are either equal or having fixed ratios, the abbreviations $S(\pi,\pi)(L_1)$ and 
$S(\pi,\pi)$ will be used for $S(\pi,\pi,L_1,L_2,L_3)$ if no confusions may 
occur.
\bibitem{beta}
The values of $\beta$ used for all the simulations with $L\le 10$ are 
$\beta J = 24.0$.
\end{thebibliography}
\end{document}